\documentclass{ws-procs975x65}
\usepackage{graphicx}
\begin{document}
\title{Future measurements of the Lense-Thirring effect in the Double Pulsar}
\author{Marcel S. Kehl $^{1}$,
			Norbert Wex$^{*1}$,
			Michael Kramer$^{1,2}$,
			Kuo Liu$^{1}$}

\address{$^1$ Max-Planck-Institut f\"ur Radioastronomie,\\
Auf dem H\"ugel 69, 53121 Bonn, Germany\\
$^2$ Jodrell Bank Centre for Astrophysics,\\
University of Manchester Oxford Road, Manchester M13 9PL, UK\\
$^*$E-mail: wex@mpifr-bonn.mpg.de\\
}

\begin{abstract}
The Double Pulsar system {PSR~J0737-3039A/B} has proven to be an excellent laboratory for high precision tests of general relativity.
With additional years of timing measurements and new telescopes like the Square Kilometre Array (SKA), the precision of these tests will increase and new effects like the Lense-Thirring precession of the orbit will become measurable.
Here, we discuss the prospects of measuring the Lense-Thirring effect and thereby 
constraining the equations of state at supra-nuclear densities
in neutron stars using the Double Pulsar.
\end{abstract}

\keywords{Tests of General Relativity; Pulsar Timing; Double Pulsar}
\bodymatter

\section{Introduction: The Double Pulsar}
The Double Pulsar discovered in 2003 (see~Ref.~\refcite{Burgay:2003jj}~and~\refcite{Lyne:2004cj}) is the only known binary system in which both neutron stars have been observed as radio pulsars.
The system consists of a mildly recycled ($P_{\rm A} = 22.7 \, \rm ms$) millisecond pulsar 
called {PSR~J0737$-$3039A} (pulsar A) and a young slower rotating pulsar 
($P_{\rm B} = 2.8 \, \rm s$) named {PSR~J0737$-$3039B} (pulsar B).
Both are orbiting each other in a close orbit with a remarkable short orbital period of only 
$P_{\rm b} = 147 \, \rm min$.
The timing observations of both pulsars yield accurate information of the system which has been used for some of the most advanced precision tests of general relativity.\cite{Kramer:2006nb}$^{,}$\cite{Kramer:2009zza}

\noindent
Due to the edge-on orientation of the Double Pulsar system there is a $\sim 30$ 
second eclipse of pulsar A 
when it passes behind the plasma-filled magnetosphere of pulsar B.\cite{Burgay:2003jj}$^{,}$\cite{Lyne:2004cj}
Based on models of this eclipse the relativistic precession of pulsar B's spin axis around the total angular moment has been measured.\cite{Breton:2008xy}
Due to the relativistic precession the beam of pulsar B moved out of our line-of-sight and has not been visible any more since 2008.\cite{Perera:2010sp}
This general relativity effect was predicted shortly after the discovery of the Hulse-Taylor pulsar.\cite{1974CRASM.279..971B} 
In contrast, pulsar A has a very stable profile (see Ref.~\refcite{Manchester:2005ev}), indicating that the spin vector is well aligned with the total angular momentum vector.\cite{Ferdman:2007nu}$^{,}$\cite{Ferdman:2013xia}
This accurate alignment makes it impossible to measure the precession of pulsar A's spin-vector around the total angular momentum vector due to relativistic spin-orbit coupling.
However, the spin-orbit coupling also yields a contribution to the total advance of periastron, $\dot{\omega}_{\rm tot} =\dot{\omega}_{1+2\, \rm PN} + \dot{\omega}_{\rm LT}$, known as the Lense-Thirring precession, $\dot{\omega}_{\rm LT}$.\cite{Barker:1975}$^{,}$\cite{Damour:1988mr}
For the Double Pulsar the Lense-Thirring contribution of pulsar B is negligible, since it spins more than 100 times slower than pulsar A.
The Lense-Thirring effect is in principal measurable if we are able to  extract the contribution of the Lense-Thirring precession $\dot{\omega}_{\rm LT}$ from the total advance of periastron 
$\dot{\omega}_{\rm tot}^{\rm obs}$ measured by timing observations. 
This can be done by estimating the rate of advance of periastron without spin effects, $\dot{\omega}_{1+2\, \rm PN}$, independently based on two other post-Keplerian (PK) parameters. 
The most promising PK parameters in the Double Pulsar are the shape of the Shapiro delay, $s$, and the orbital period decay, $\dot{P}_{\rm b}$, due to gravitational wave dumping.\cite{Kramer:2009zza}
The difference between the measured $\dot{\omega}_{\rm tot}^{\rm obs}$ and the independently estimated value for $\dot{\omega}_{1+2\, \rm PN}$ should then correspond to the contribution of the Lense-Thirring precession.

\section{Measuring the Lense-Thirring effect with future timing of the  Double Pulsar}
\begin{figure}[htb!]
\begin{center}
\includegraphics[width=4.3in]{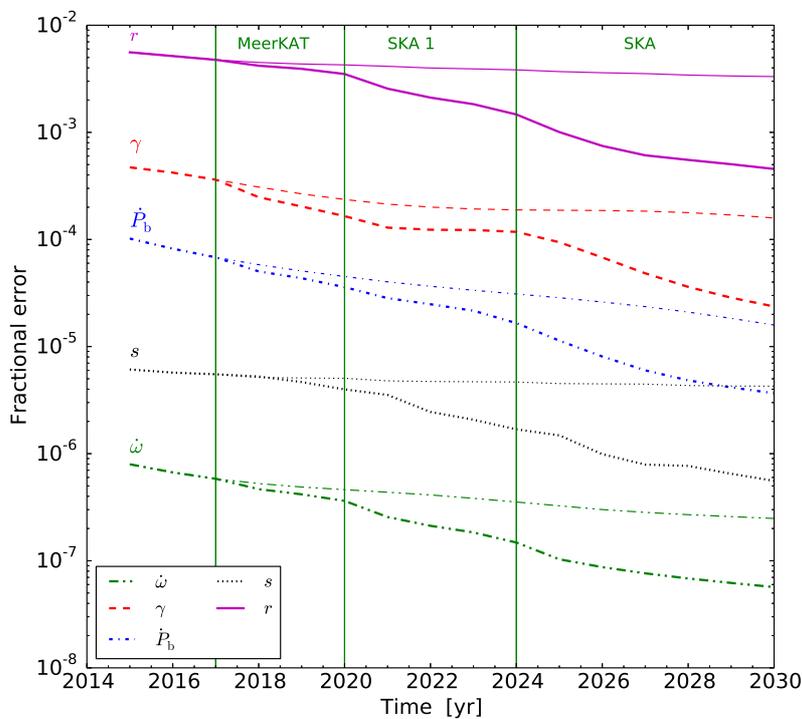}
\end{center}
\caption{Expected evolution of the fractional timing precision error of the post-Keplerian parameters in the Double Pulsar based on our simulations. The bold lines correspond to simulations with future telescopes, whereas the thin lines indicate the evolution without future telescopes.}
\label{fig:frac-error}
\end{figure}
\noindent
We performed mock data simulations of the pulsar timing data from {PSR J0737-3039A} 
in order to estimate our capability of measuring the Lense-Thirring precession 
$\dot{\omega}_{\rm LT}$ in the Double Pulsar.
In our simulations, we incorporate the contribution of the Lense-Thirring effect 
and a noise model which accounts for radiometer noise due to the system temperature and jitter noise by stochastic variability in the pulsar signal.\cite{Liu:2011}
We also consider the improvements of the system sensitivity achieved by future radio telescopes such as MeerKAT and SKA. 
In a first step, we reproduced the current timing precision of the system and then generated fake data up to the year 2030. 
The simulated timing data are analysed with the pulsar timing software TEMPO\footnote{\url{http://tempo.sourceforge.net/}}.
Based on the fitted timing parameters (see~Fig.~\ref{fig:frac-error}), specially the post-Keplerian parameters $\dot{P}_{\rm b}, s, \dot{\omega}$, we performed Monte-Carlo simulations to extract the Lense-Thirring precession $\dot{\omega}_{\rm LT}$ 
and its uncertainty for each year. 
The results of our simulations (see~Fig.~\ref{fig:Error}) indicate that we will be able to obtain 
a $3\sigma$-measurement of the Lense-Thirring effect in the 
Double Pulsar after only a few years of SKA~1 timing. 
This will yield a new quantitative test of the Lense-Thirring effect 
in the quasi-stationary strong field regime of gravity.

\noindent
Once we have a significant measurement of the Lense-Thirring precession, we can follow a new approach.
The special geometry of the Double Pulsar system allows us to estimate the 
moment of inertia of pulsar A (see Ref.~\refcite{Damour:1988mr}) based on a measurement on the Lense-Thirring precession. 
A measurement of the moment of inertia can be directly used to restrict the equations of state (EoS) for nuclear matter at ultra high densities.\cite{Lattimer:2004nj}
Based on our mock data simulations we find that we will be able to distinguish between several different equations of state after a few years of full SKA timing~(2030), and measure the moment of inertia with an accuracy below $10\%$ by this time.
Neglecting the increased gain of future telescopes, i.e. MeerKAT and SKA, in our simulations demonstrates that instrumental improvements (especially the SKA) 
are crucial for reaching the precision needed to measure the moment of inertia with 
less than $10\%$ precision and to distinguish between equations of state till this time (see~Fig.~\ref{fig:Error}).

\section{External Effects}
\begin{figure}[ht!]
\begin{center}
\includegraphics[width=4.2in]{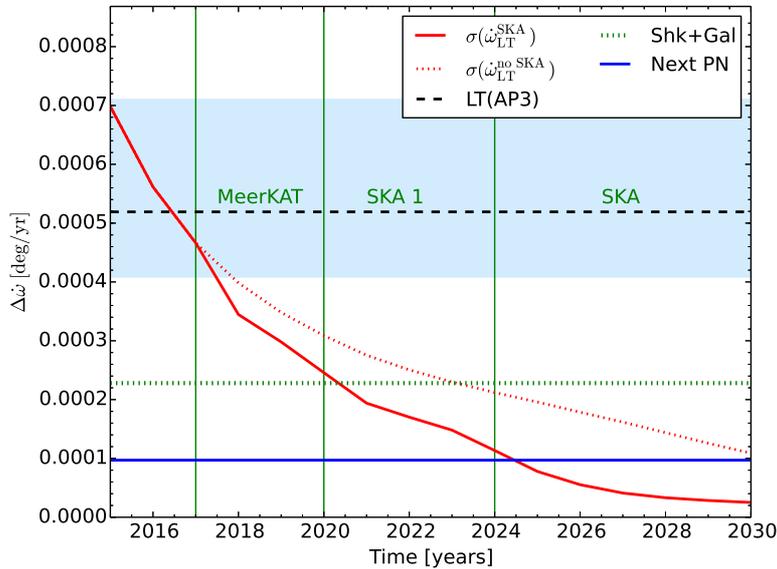}
\end{center}
\caption{Improving sensitivity, $\Delta \omega_{\rm LT}$ (red curves), for measuring the Lense-Thirring effect based on mock data simulations. The blue shaded area indicates the expected magnitude of the Lense-Thirring effect for different equations of state (lower edge: WFF1 -- upper edge:~MS0). Vertical lines indicate different telescope stages. 
The uncertainty of the observed rate of orbital period change due to the galactic acceleration and relative motion of the Double Pulsar is given by the green horizontal line.}
\label{fig:Error}
\end{figure}
Our ability to measure the Lense-Thirring effect is currently limited by the precision of  the orbital period decay $\dot{P}_{\rm b}$.
In the Double Pulsar, we can directly link the uncertainty of the orbital period
decay, $\Delta \dot{P}_{\rm b}$, to the uncertainty of our Lense-Thirring measurement, $\Delta \dot{\omega}_{\rm LT}$, as $\Delta \dot{P}_{\rm b} (\Delta \dot{\omega}_{\rm LT} ) \approx 1.82 \cdot 10^{-13} \, \Delta \dot{\omega}_{\rm LT} [\rm deg/yr]$. 

\begin{figure}[ht!]
\begin{center}
\includegraphics[width=5.1in]{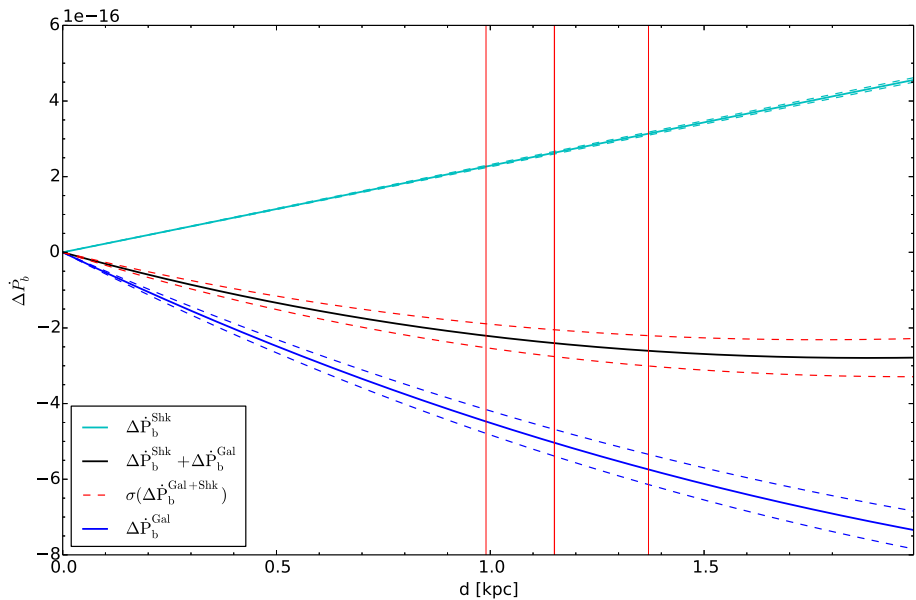}
\end{center}
\caption{Contributions of the Shklovskii effect and the Galactic acceleration on the observed orbital period derivative $\dot{P_{b}}$ as function of the distance. The central vertical red line indicates the currently best known distance measurement. The thin red lines show the corresponding $1\sigma$-error.}
\label{fig:Shk-Gal}
\end{figure}

\noindent
The measured orbital period decay $\dot{P}_{\rm b}$ is contaminated by  external effects which have to be subtracted to obtain the intrinsic value.
Today the most prominent contributions are the Shklovskii effect (see~Ref.~\refcite{Shklovskii:1970}) and the Galactic acceleration (see~Ref.~\refcite{Damour:1990wz}), which both depend on the distance of the system.
Using current Galactic models (see Ref.~\refcite{Reid:2014boa}) and distance measurements (see Ref.~\refcite{Deller:2009dk}), we find that the sum of these two effects only weakly depends on the actual distance of the Double Pulsar (see~Fig.~\ref{fig:Shk-Gal}). 
However, an improvement of our Galactic model is essential for measuring the 
Lense-Thirring effect with more than $2\sigma$-precision (see~Fig.~\ref{fig:Error}). 
Such an improvement is expected with future observations of the GAIA satellite\footnote{\url{http://sci.esa.int/gaia/}} and ongoing observations of Galactic masers (see~Ref.~\refcite{Reid:2014boa}). 
Moreover, we have to include next-to-leading order post-Newtonian contributions of the 
orbital period decay for the first time in order 
to achieve a precision of below $\sim 20\%$.
The orbital period as an periodic quantity is affected by the Doppler effect 
due to the relative movement of the system and the observer.
Therefore, relative accelerations also affect our capability of measuring the 
orbital period decay. 
In addition to the Galactic acceleration, further accelerations can  
be introduced by stars, molecular clouds and other local masses. 
Our first analysis of the molecular gas in the environment 
(see~Ref.~\refcite{Dame:2000sp}) of the Double Pulsar, 
let us conclude that the system is located at a favourable location within the Milky Way.
However, future analysis based on observations are needed to understand the local accelerations in more detail.
Based on our current knowledge and investigations, we expect that these acceleration effects do not limit our capability of measuring the Lense-Thirring effect with $\sim 10\%$ sensitivity. 

\section{Discussion}
As a result of our mock data simulations for the Double Pulsar, we found that the Lense-Thirring effect will be measurable with $3\sigma$-accuracy after a few years of SKA~1 timing.
This measurement will yield a new quantitative way of testing the Lense-Thirring effect in the quasi-stationary strong field regime of gravity.
The considerable increased timing precision with full SKA,  
will theoretically enable us to measure the Lense-Thirring effect with 
an accuracy below $10\%$, given that we can control the external effects.
Such a precise measurement can be used to restrict the moment of inertia, 
the radius as well as the equation 
of state for the neutron star {PSR~J0737$-$3039A}.
This yields a different approach from those based on X-ray observations and built upon certain model assumptions (see~e.g.~Ref.~\refcite{Oezel:2016}). 

\noindent
However, further investigations on the impacts of short time variations of the electron density between the Double Pulsar and Earth as well as  detailed analyses on the role of non-Gaussian red noise contributions in the Double Pulsar are needed. 

\bibliographystyle{unsrt}

\end{document}